\documentclass[]{article}
\pdfoutput=1
\usepackage[utf8]{inputenc}

\usepackage{arxiv}
\usepackage{booktabs} 
\usepackage{listings}
\usepackage{graphicx}
\usepackage{hyperref}
\usepackage{amsmath}
\usepackage{amssymb}

\usepackage[T1]{fontenc}
\usepackage{subcaption} 
\usepackage{verbatim} 

\author{
Sarah Bird\\
Mozilla\\
\texttt{sbird@mozilla.com}
}
   
\title{\huge Tracking measurement obfuscations from sourceURL}
\date{\today}

\begin{document}
\maketitle

\begin{abstract}
Tracking scripts can use the \texttt{sourceURL} directive to mask their origin from developer tools and tools that use the same JS call stack and network stack information.
Firefox and Chromium appear to be affected.
Firefox 78 now includes a preference to disable this behavior.
This short paper describes the effect when using the OpenWPM measurement platform along with details of discovery.

\end{abstract}


\section{Background}
\label{background}

\subsection{Tracking protection backed by OpenWPM}

Firefox's tracking protection is powered by the Disconnect list \cite{disconnectDisconnectmeDisconnecttrackingprotection2020} an open source list of trackers that is used by a number of tracking protection softwares.
Disconnect is an independent organization, but they collaborate with Mozilla. 
In particular, Mozilla provides Disconnect with lists of suspected fingerprinters for them to review and add to the list \cite{MozillaSecurityResearchb}.

Mozilla's method for building a list of suspected fingerprinters largely builds on Englehardt's doctoral work \cite{englehardtAutomatedDiscoveryPrivacy} on the automated discovery of privacy violations, in particular the 2016 paper \textit{Online Tracking: A 1-million-site measurement and analysis} \cite{englehardtOnlineTracking1millionsite2016}.
The method uses the open source framework OpenWPM \cite{MozillaOpenWPM2020} to instrument the Firefox web browser enabling measurement of JavaScript API access, and network requests in particular. 

To detect fingerprinting, heuristics are used as outlined in Englehardt's \cite{englehardtOnlineTracking1millionsite2016} and Das et al. \cite{dasWebSixthSense2018}. 
These heuristics label a script as audio, canvas, webrtc, or font fingerprinting if OpenWPM has detected them accessing a specific set of APIs in specific conditions. 
For example, to detect font fingerprinting, the heuristic looks for a script which draws a small amount of text to canvas over 50 times with 50 different scripts.

Upon detecting scripts that engage in fingerprinting, the script URLs are collected to identify domains which may need to be blocked due to fingerprinting activity.

\subsection{Details of OpenWPM instrumentation}

To detect fingerprinting scripts, the OpenWPM extension instruments a set of JS APIs known to be associated with fingerprinting.
It does this by overwriting methods on \texttt{Object} properties and functions so that when those functions are called or properties are get or set a logging step kicks in before providing the original value to the script calling the instrumented item.
During the logging step, a set of metadata is collected for later analysis including the value that was \texttt{set} or \texttt{get}, any arguments that were sent to a function, etc. 
Crucially, information about the calling script is captured. 
This is done by throwing an error and capturing the call stack. 
The call stack contains the information about what code was executed while arriving at the instrumented call.
This call stack information is the same that is available to website developers in the browser developer tools \cite{mdncontributorsWhatAreBrowser}.
OpenWPM parses the call stack to generate a \texttt{callContext} which contains the originating script along with the function the code was in when the the call was made.

\subsection{\texttt{sourceURL} directive}

To facilitate debugging of code, browsers implement directives, or pragmas. 
As described on MDN \cite{mdncontributorsDebugEvalSources} and Chrome DevTools docs \cite{kearneyMapPreprocessedCode}, the \texttt{sourceURL} directive can be used to give code that is evaluated dynamically a name.
This enables software developers to be able to see more easily what code is executing in developer tools.

The \texttt{sourceURL} directive is related to the \texttt{sourceMappingURL} directive.
A \texttt{sourceMappingURL} directive relates minified and aggregated client-side code with developer's original, easy to read, source code.

The Source Map Revision 3 Proposal \cite{lenzSourceMapRevision2011} outlines the specifics of how the \texttt{sourceMappingURL} directive should work. 
The spec mentions \texttt{sourceURL} as a way to determine source origin when "generated source is not associated with a script element that has a 'src' attribute".

\section{\texttt{sourceURL} obfuscation}

\subsection{Discovery}

An OpenWPM crawl was performed on the Alexa top 100k sites in February 2020. 
After applying heuristics, 3,134 fingerprinting scripts from 2,203 fingerprinting script domains were found on 5,106 different visited domains.

The original analysis was trying to match fingerprinting script discoveries with requests in the OpenWPM requests table.
Because javascript is instrumented separately from the requests table a manual join must be performed.
When attempting to join on the \texttt{script\_url} from the javascript table to the \texttt{a} in the requests table, a number of scripts did not match.

Further investigation narrowed this to a handful of \texttt{script\_url}s. 
The four most common were an empty string \texttt{''}, \texttt{(program):2}, \texttt{dna.min.js}, \texttt{dna-persist.min.js}. 

The empty string can be expected to occur when code is inline with the main html page and it was not looked at further.
The last three stood out as they are not full URLs.
For these, the sites where these scripts were observed were manually loaded and developer tools used to inspect the incoming requests and scripts on the page.

In discussing the mismatch with the OpenWPM developer team, it was identified that \texttt{sourceURL} was the common thread that all the scripts had set and was causing the instrumentation mismatch.

\subsection{Reporting}

A bug was filed with OpenWPM \cite{birdSourceURLObfuscatesScript} at the end of March 2020, initially hoping that OpenWPM would need to change a preference in order to access the unmasked \texttt{script\_url}.
After further investigation it appeared that no such preference was available and a bug was filed with Firefox \cite{bird1628853ExposeOption}.
When triage was unable to view the issue on an example site, a test site was set up to demonstrate the problem in isolation. 
The test page is available at \url{http://www.sarahbird.org/tests/sourceURL/}.

During triage it was also established that along with the JavaScript callstack, the network call stack was affected.
The test page reflects both these issues.

During triage it was decided that a pref should be added to block the use of not only the \texttt{sourceURL} directive, but also \texttt{sourceMappingURL}. 
Although \texttt{sourceMappingURL} had not been identified as a problem, it is logical to want to turn all these behaviors off together.

\subsection{Fix}

In early May 2020, Smyth added the preference \texttt{javascript.options.source\_pragmas} to Firefox 78 \cite{smythBug1628853Expose}. 
The preference can be changed from the default \texttt{true} to \texttt{false} in \texttt{about:config} to disable \texttt{sourceURL} and \texttt{sourceMappingURL} directives.

\subsection{Other Browser implementations}
\label{other-browser}

Having narrowed the source of the problem and created a test page, the test page was tested with Chromium and Safari browsers.

Chromium appears to exhibit the same behavior as Firefox and the \texttt{sourceURL} is masking the identity on the JS stack and network stack and cannot be turned off. 

Safari appears to not suffer from this problem.

\section{Future Work}
\label{future-work}

A number of elements related to this issue have not yet been investigated: 

\begin{itemize}
    \item In finalizing the bug \cite{bird1628853ExposeOption} to describe the issue, it was decided that the related \texttt{sourceMappingURL} pragma should also be included in the fix. However, \texttt{sourceMappingURL} has not been investigated with this work or identified as being used for obfuscation in the wild.
    \item All identified fingerprinting scripts for the crawl were identified using heuristics for canvas, font, audio, and webrtc fingerprinting. These heuristics do not identify all fingerprinting scripts.
    \item A large set of scripts downloaded by the crawl were reviewed for presence of sourceURL. Of note were scripts from PerimeterX, a company on the Disconnect list. These scripts contain code that references, and potentially manipulates the \texttt{sourceURL}, but this has not been investigated yet.
\end{itemize}

\bibliographystyle{plainurl}
\bibliography{bibliography.bib}

\end{document}